\documentclass[oneside,english]{amsart}
\usepackage[T1]{fontenc}
\usepackage[latin9]{inputenc}
\usepackage{float}
\usepackage{amsthm}
\usepackage{amsbsy}
\usepackage{amstext}
\usepackage{amssymb}
\usepackage{graphicx}
\PassOptionsToPackage{normalem}{ulem}
\usepackage{ulem}

\makeatletter

\providecommand{\tabularnewline}{\\}

\numberwithin{equation}{section}
\numberwithin{figure}{section}
\theoremstyle{plain}
\newtheorem{thm}{\protect\theoremname}
  \theoremstyle{definition}
  \newtheorem{defn}[thm]{\protect\definitionname}

\makeatother

\usepackage{babel}
  \providecommand{\definitionname}{Definition}
\providecommand{\theoremname}{Theorem}

\begin{document}

\title{Compressive Sampling Using EM Algorithm}

\author{Atanu Kumar Ghosh,\ Arnab Chakraborty}

\address{Applied Statistics Unit, Indian Statistical Institute}
\begin{abstract}
Conventional approaches of sampling signals follow the celebrated
theorem of Nyquist and Shannon. Compressive sampling, introduced by
Donoho, Romberg and Tao, is a new paradigm that goes against the conventional
methods in data acquisition and provides a way of recovering signals
using fewer samples than the traditional methods use. Here we suggest
an alternative way of reconstructing the original signals in compressive
sampling using EM algorithm. We first propose a naive approach which
has certain computational difficulties and subsequently modify it
to a new approach which performs better than the conventional methods
of compressive sampling. The comparison of the different approaches
and the performance of the new approach has been studied using simulated
data. 
\end{abstract}
\maketitle

\section{Introduction}

In recent years there has been a huge explosion in the variety of
sensors and the dimensionality of the data produced by these sensors
and this has been in a large number of applications ranging from imaging
to other scientific applications.The total amount of data produced
by the sensors is much more than the available storage. So we often
need to store a subset of the data. We want to reconstruct the entire
data from it. The famous Nyquist-Shannon sampling theorem {[}5{]}
tells us that if we can sample a signal at twice its highest frequency
we can recover it exactly. In applications this often results in too
many samples which must be compressed in order to store or transmit.
An alternative is compressive sampling (CS) which provides a more
general data acquisition protocol by reducing the signal directly
into a compressed representation by taking linear combinations. In
this paper we present a brief of the conventional approach of compressive
sampling and propose a new approach that makes use of the EM algorithm
to reconstruct the entire signal from the compressed signals.

\section{setup}

When a signal is sparse in some basis , a few well chosen observations
suffice to reconstruct the most significant nonzero components.

\paragraph*{Consider a signal $\mathbf{x}$ represented in terms of a basis expansion
as}

\[
\mathbf{x}={\displaystyle \sum_{i=1}^{n}s_{i}\psi_{i}=\mathbf{\psi s}}
\]
The basis $\mathbf{s}$ is such that only $k<<n$ coefficients $\psi_{i}$
have significant magnitude. Many natural and artificial signals are
sparse in the sense that there exists a basis where the above representation
has just a few large coefficients and other small coefficients. As
an example natural images are likely to be compressible in discrete
cosine transform(DCT) and wavelet bases {[}1{]}. In general we do
not know apriori which coefficients are significant. The data collected
by a measurement system consists of some linear combinations of the
signals 

\[
\mathbf{y}=\phi\mathbf{x}+\mathbf{e=\phi\psi s+e=As+e}
\]
where $A=\phi\psi$ is a measurement matrix (also called sensing matrix)
which is chosen by the statistician. The measurement process is non-adaptive
as $\phi$(and hence $A)$ does not depend in any way on the signal
$\mathbf{x}$. $\mathbf{e}$ is the error which is assumed to be bounded
or bounded with high probability.

\medskip{}

\paragraph*{Our aim here is to :}
\begin{itemize}
\item design a stable measurement matrix that preserves the information
in any $k$-sparse signal during the dimensionality reduction from
$\mathbb{R}^{n}$ to $\mathbb{R}^{m}$.
\item design a reconstruction algorithm to recover the original data $\mathbf{x}$
from the measurements $\mathbf{y}$.
\end{itemize}
\medskip{}

\medskip{}

\paragraph*{We note that the recovery algorithm addresses the problem of solving
for $\mathbf{x}$ when the number of unknowns (i.e. $n$) is much
larger than the number of observations (i.e. $m$) . In general this
is an ill-posed problem but CS theory provides a condition on $\phi$
which allows accurate estimation.}

One such popularly used property is Restricted Isometry Property (RIP)
{[}2{]}. 
\begin{defn}
The matrix $A$ satisfies the restricted isometry property of order
$k$ with parameters $\delta_{k}\in[0,1)$ if 

\[
(1-\delta_{k})\parallel\theta\parallel_{2}^{2}\leq\parallel A\theta\parallel_{2}^{2}\leq(1+\delta_{k})\parallel\theta\parallel_{2}^{2}
\]
holds simultaneously for all sparse vectors $\theta$ having no more
than $k$ nonzero entries. Matrices with this property are denoted
by RIP($K,\delta_{k})$
\end{defn}

\section{Conventional Approach}

\paragraph*{The following theorem shows that matrices satisfying RIP will yield
accurate estimates of $\mathbf{x}$ with the help of recovery algorithms.}
\begin{thm}
Let $A$ be a matrix satisfying RIP$(2k,\delta_{2k})$ with $\delta_{2k}<\sqrt{2}-1$
and let $\mathbf{y}=A\mathbf{s}+\mathbf{e}$ be a vector of noisy
observations , where $\parallel\mathbf{e}\parallel_{2}\leq\epsilon$.
Let $\mathbf{s}_{k}$ be the best $k$-sparse approximation of $\mathbf{s}$
, that is , $\mathbf{s_{k}}$ is the approximation obtained by keeping
the $k$ largest entries of $s$ and setting others to zero. Then
the estimate

\begin{align}
\hat{\mathbf{s}} & =\arg\min_{\mathbf{s}\in\mathbb{R}^{n}}\parallel\mathbf{s}\parallel_{1}\;\textrm{subject to }\parallel\mathbf{y}-A\mathbf{s}\parallel_{2}\leq\epsilon\label{eq:equation3.1}
\end{align}
obeys
\begin{equation}
\parallel\mathbf{s-\hat{s}}\parallel_{2}\leq C_{1,k}\epsilon+C_{2,k}\frac{\parallel\mathbf{s-}\mathbf{s}_{k}\parallel_{1}}{\sqrt{k}}
\end{equation}
where $C_{1,k}$ and $C_{2,k}$ are constants depending on $k$ but
not on $n$ or $m$.
\end{thm}

\paragraph*{The reconstruction in (3.1) is equivalent to }

\begin{equation}
\hat{\mathbf{s}}=\arg\min_{\mathbf{s}\in\mathbb{R}^{n}}\frac{1}{2}\parallel\mathbf{y}-A\mathbf{s}\parallel_{2}^{2}+\lambda\parallel\mathbf{s}\parallel_{1}\;\textrm{ and }\hat{\mathbf{x}}=\psi\hat{\mathbf{s}}\label{eq:equation3.3}
\end{equation}
where $\lambda>0$ is a regularization parameter which depends on
$\epsilon$.

\section{A Naive Approach\label{sec:A-Naive-Approach}}

In this approach we apply EM algorithm for the reconstruction of the
signal. Since we observe some linear combinations of the signals instead
of the entire signals we can treat the observed linear combinations
as our observed data and the entire signals as the complete data which
is unobserved. Hence we apply EM algorithm as a most natural tool
of missing data analysis to reconstruct the data. Here we assume that
data are coming from a population with mean $\boldsymbol{\mathbf{\mu}}$
and that $\boldsymbol{\mathbf{\mu}}$ is sparse (w.r.t some basis).
Without loss of generality we assume that $\boldsymbol{\mathbf{\mu}}$
is sparse with respect to euclidean basis.We assume that at most $k$
elements of \textbf{$\boldsymbol{\mathbf{\mu}}$} is nonzero.

\paragraph*{Let us assume that the parent population is normal viz. $N(\mathbf{\boldsymbol{\mathbf{\mu}}},\sigma^{2}I_{n})$}

\paragraph*{Then we have the signal as 
\[
\mathbf{x=\boldsymbol{\mathbf{\mu}}+\boldsymbol{\epsilon}}
\]
 where $\mathbf{\boldsymbol{\epsilon}}\sim N(0,I_{n})$}

Then with the help of the sensing matrix we have the observed data
as

\[
\mathbf{y}=\phi\mathbf{x}=\phi(\mathbf{\mu+\epsilon})=\phi\mathbf{\boldsymbol{\mathbf{\mu}}+e}
\]
 where $\mathbf{e=\phi\boldsymbol{\epsilon}}$

\paragraph*{Thus unlike the conventional approach here we assume that the signals
themselves are subject to error and consequently the observed combinations
of the signals are also subject to error. Here we try to reconstruct
the unobserved true signals which are free from error. }

\medskip{}

\medskip{}

We then treat $\mathbf{x}$ as the complete data and $\mathbf{y}$
as the observed data and try to estimate $\mathbf{\mathbf{\boldsymbol{\mathbf{\mu}}}}$
from the observed data using EM algorithm .Thus we have\medskip{}
\[
\mathbf{\boldsymbol{\mathbf{\mu}}}=(\mu_{1},\mu_{2}...\mu_{n})^{'}
\]

\[
\mathbf{x}\sim N_{n}(\boldsymbol{\mathbf{\mu}},\sigma^{2}I_{n})\;\textrm{:Complete data}
\]
 
\[
\mathbf{y}=\phi\mathbf{x}\sim N_{m}(\phi\boldsymbol{\mathbf{\mu}},\sigma^{2}\phi\phi^{'})\;\textrm{:Observed data}
\]

\medskip{}

\paragraph*{The complete data likelihood is given by
\[
f(\mathbf{x)=}\frac{1}{(\sigma\sqrt{2\pi})^{n}}e^{-\frac{1}{2\sigma^{2}}\mathbf{(x-\boldsymbol{\mathbf{\mu}})^{'}(x-\boldsymbol{\mathbf{\mu}})}},\mathbf{x}\in\mathbb{R}^{n},\mathbf{\boldsymbol{\mathbf{\mu}}}\in\mathbb{R}^{n},\sigma>0
\]
The conditional distribution of the complete data given the observed
data is }

\[
\mathbf{x|y,\boldsymbol{\mathbf{\mu}}\sim N_{n}(\boldsymbol{\mathbf{\mu}}+\phi^{'}(\phi\phi^{'})^{-1}(y-\phi\boldsymbol{\mathbf{\mu}}),\;\sigma^{2}(I_{n}-\phi^{'}(\phi\phi^{'})^{-1}\phi}))
\]

\paragraph*{After $t$ iterations in EM algorithm we have,\medskip{}
}

\medskip{}

\begin{itemize}
\item \textbf{\uline{E Step}}:We compute the expected complete data log-likelihood
w.r.t the conditional distribution of $\mathbf{x|y,\boldsymbol{\mathbf{\mu}}}^{(t)}$.Now
\[
\ell(\boldsymbol{\mathbf{\mu}})=\ln(f(\mathbf{x}))=constant-\frac{1}{2}(\mathbf{x-\boldsymbol{\mathbf{\mu}})^{'}(x-\boldsymbol{\mathbf{\mu}}})
\]
\[
\Rightarrow\ell(\boldsymbol{\mathbf{\mu}})=constant-\frac{1}{2}\sum_{i=1}^{n}(x_{i}-\mu_{i})^{2}
\]
Also 
\[
\mathbf{x|y,\boldsymbol{\mathbf{\mu}}^{(t)}\sim N_{n}(\mu^{(t)}+\phi^{'}(\phi\phi^{'})^{-1}(y-\phi\boldsymbol{\mathbf{\mu}}^{(t)}),\;\sigma^{2}(I_{n}-\phi^{'}(\phi\phi^{'})^{-1}\phi}))
\]
Define
\[
Q(\mathbf{\boldsymbol{\mathbf{\mu}}})=E(\ell(\boldsymbol{\mathbf{\mu}})|y,\boldsymbol{\mathbf{\mu}}^{(t)})
\]
\medskip{}
\medskip{}

\item \textbf{\uline{M Step}}:Here we try to maximize $Q(\mathbf{\boldsymbol{\mathbf{\mu}}})$
with respect to $\boldsymbol{\mathbf{\mu}}$.We know that $\mathbf{\boldsymbol{\mathbf{\mu}}}$
is sparse i.e. some of the $\mu_{i}$ are zero. So we need to maximize
$Q(\boldsymbol{\mathbf{\mu}})$ w.r.t. $\boldsymbol{\mathbf{\mu}}$
belonging to a subset 
\[
S=\{\boldsymbol{\mathbf{\mu}}:\textrm{at most \ensuremath{k}elements of \ensuremath{\boldsymbol{\mathbf{\mu}}}\ are nonzero}\}
\]
Thus we find 
\[
\arg\max_{\boldsymbol{\mu}\in S}Q(\mu)
\]

\end{itemize}
\medskip{}

\paragraph*{For this we note that $S=\cup_{i=1}^{{n \choose k}}S_{i}$ where
\[
S_{i}=\{\boldsymbol{\mu}:\textrm{at most \ensuremath{i}specific elements of \ensuremath{\boldsymbol{\mu}}\ are nonzero}\}
\]
We then find 
\[
\arg\max_{\boldsymbol{\mu}\in S_{i}}Q(\boldsymbol{\mu})
\]
 for each $i$ and call the estimate as 
\[
\mathbf{\hat{\mu}}^{(t+1)}(S_{i})=(\hat{\mu}_{1}^{(t+1)}(S_{i}),\hat{\mu}_{2}^{(t+1)}(S_{i})...\hat{,\mu}_{k}^{(t+1)}(S_{i}))^{'}
\]
Now the \textmd{${\displaystyle \arg\max_{\boldsymbol{\mu}\in S_{i}}}Q(\boldsymbol{\mu})$
is found out in the following way:}\medskip{}
}

\paragraph*{Setting $\frac{\partial}{\partial\mu_{j}}Q(\boldsymbol{\mu})=0$
for those $j$ such that $\mu_{j}\neq0$ we find that
\[
\hat{\mu}_{j}^{(t+1)}(S_{i})=\mu_{j}^{(t)}+\alpha_{j}+\beta_{j}
\]
 where $\alpha_{j}=\textrm{\ensuremath{j^{th}}element of}\:\phi^{'}(\phi\phi^{'})^{-1}y$
and $\beta_{j}=\textrm{\ensuremath{j^{th}}element of}\:\phi^{'}(\phi\phi^{'})^{-1}\phi\boldsymbol{\mu}^{(t)}$.\medskip{}
}

\paragraph*{Then we choose the $\mathbf{\hat{\boldsymbol{\mu}}}^{(t+1)}(S_{i})$
for which $Q(\mathbf{\hat{\boldsymbol{\mu}}}^{(t+1)}(S_{i}))$ is
maximum as the new estimate of $\mathbf{\boldsymbol{\mu}}$ at $(t+1)^{th}$
iteration.Thus the estimate of $\boldsymbol{\mu}$ is
\[
\hat{\boldsymbol{\mu}}^{(t+1)}=\hat{\boldsymbol{\mu}}^{(t+1)}(S_{i})
\]
such that
\[
Q(\mathbf{\hat{\boldsymbol{\mu}}}^{(t+1)}(S_{i}))\geq Q(\mathbf{\hat{\boldsymbol{\mu}}}^{(t+1)}(S_{j}))\:\forall j\neq i
\]
We iterate until convergence.}

\section{new approach}

The new approach discussed in the previous section requires the maximization
of $Q(\boldsymbol{\mu})$ over ${n \choose k}$ subspaces and then
choose the one for which it is maximum at the M step of each EM iteration.
This is computationally expensive and practically impossible to implement
for large $n$. Hence we suggest an alternative way which instead
of maximization over ${n \choose k}$ subspaces in each EM iteration
identifies a particular subspace where $\boldsymbol{\mu}$ is most
likely to belong , and then finds the maximum over that subspace in
each M step.

\paragraph*{Let $S_{\boldsymbol{\mu}}$ be the subspace where $\boldsymbol{\mu}$
lies , that is 
\[
S_{\boldsymbol{\mu}}=\{(x_{1},x_{2},\ldots,x_{n})\in\mathbb{R}^{n}:\:\forall i\;\mu_{i}=0\:\Rightarrow x_{i}=0\}
\]
 We note that if we find the unrestricted maximizer of $Q(\boldsymbol{\mu})$
in each M step of the EM algorithm (henceforth call unrestricted EM
) , that is if we find 
\[
\hat{\boldsymbol{\mu}}^{un}=\arg\max_{\boldsymbol{\mu}\in\mathbb{R}}Q(\boldsymbol{\mu})
\]
then the unrestricted EM estimate $\hat{\boldsymbol{\mu}}^{un}$ should
lie close to $S_{\boldsymbol{\mu}}$. Hence the unrestricted estimate
should provide an indication of the subspace in which the original
parameter lies. Hence we find which components of $\hat{\boldsymbol{\mu}}^{un}$
are significant so that we can take the other insignificant components
to be zero and take the corresponding subspace thus formed to be the
one in which our estimate should lie. We test which components of
$\hat{\boldsymbol{\mu}}^{un}$ are significantly different from zero.}

Now for the unrestricted EM algorithm the estimate of $\boldsymbol{\mu}$
should converge to the maximizer of the observed log-likelihood. The
observed log-likelihood is 
\[
\ell_{obs}(\boldsymbol{\mu})=constant-\frac{1}{2}(\mathbf{y}-\phi\boldsymbol{\mu})^{'}(\sigma^{2}\phi\phi^{'})^{-1}(\mathbf{y}-\phi\boldsymbol{\mu})
\]
Setting $\frac{\partial}{\partial\boldsymbol{\mu}}\ell_{obs}(\boldsymbol{\mu})=0$
we get 
\begin{equation}
(\phi^{'}V^{-1}\phi)\boldsymbol{\mu}=\phi^{'}V^{-1}\phi\mathbf{y}\label{eq:myequation1}
\end{equation}
where $V=\phi\phi^{'}$.

\paragraph*{The above equation $\eqref{eq:myequation1}$ does not have a unique
solution as $rank[(\phi^{'}V^{-1}\phi)_{n\times n}]=m\ll n$. Hence
the observed likelihood does not have a unique maximum and our unrestricted
EM algorithm will produce many estimates of $\boldsymbol{\mu}$. Among
these many estimates we choose the sparsest solution. This is taken
care of by taking the initial estimate of $\boldsymbol{\mu}$as $\boldsymbol{0}$
in the iterative process as then the estimate will hopefully converge
to nearest solution which will be the sparest one. We will justify
this later with the help of simulation.}

\paragraph*{We have 
\[
\hat{\boldsymbol{\mu}}^{un}=(\phi^{'}V^{-1}\phi)^{+}\phi^{'}V^{-1}y=P\mathbf{y}
\]
\medskip{}
 where $P=(\phi^{'}V^{-1}\phi)^{+}\phi^{'}V^{-1}$\medskip{}
}

\paragraph*{Here we take the Moore-Penrose inverse of $(\phi^{'}V^{-1}\phi)$
as we want to find the least norm solution of $\eqref{eq:myequation1}$
. }

\paragraph*{Now 
\[
\hat{\boldsymbol{\mu}}^{un}\sim\mathbf{N_{n}(}P\phi\boldsymbol{\mu}\:,\: PVP^{'}\mathbf{)}
\]
Thus $E(\hat{\boldsymbol{\mu}}^{un})=P\phi\boldsymbol{\mu}$ and $\hat{\boldsymbol{\mu}}^{un}$
should lie close to the sparse$\boldsymbol{\mu}$. Hence $P\phi\boldsymbol{\mu}$
should be close to $\boldsymbol{\mu}$and \textmd{$\hat{\boldsymbol{\mu}}^{un}$
is used to test hypotheses regarding $\boldsymbol{\mu}$.}}

\paragraph*{We want to test $n$ hypotheses 
\[
H_{0i}:\mu_{i}=0\qquad\forall i=1(1)n
\]
Let 
\[
\hat{\boldsymbol{\mu}}^{un}=(\hat{\mu}_{1}^{un},\hat{\mu}_{2}^{un},\ldots\hat{\mu}_{n}^{un})^{'}
\]
Then the test statistics for testing $H_{0i}$ is 
\[
\tau_{i}=|\frac{\hat{\mu}_{i}^{un}}{\sqrt{s_{ii}}}|\sim N(0,1)\quad\textrm{under \ensuremath{H_{0i}}}\qquad\forall i=1(1)n
\]
where $s_{ii}=i^{th}\textrm{diagonal element of }PVP^{'}$}

\paragraph*{Thus we estimate the subspace where $\boldsymbol{\mu}$ lies as 
\[
\hat{S}_{\boldsymbol{\mu}}=\{(x_{1},x_{2},\ldots,x_{n})\in\mathbb{R}^{n}:\:\forall i\;\tau_{i}\leq z_{\alpha/2}\:\Rightarrow x_{i}=0\}
\]
With this new estimated subspace we apply our original restricted
EM algorithm as in the previous section as follows:}

\paragraph*{After $t$ iterations in EM algorithm we have,}
\begin{itemize}
\item \textbf{\uline{E step}}: Compute $Q(\mathbf{\boldsymbol{\mu}})=E(\ell(\boldsymbol{\mu})|y,\boldsymbol{\mu}^{(t)})$
\item \textbf{\uline{M-step}}: We find 
\[
\arg\max_{\boldsymbol{\mu}\in\hat{S}_{\boldsymbol{\mu}}}Q(\boldsymbol{\mu})
\]
and take the maximizer as the new estimate of $\boldsymbol{\mu}$
, that is , $\hat{\boldsymbol{\mu}}^{(t+1)}$.
\end{itemize}

\paragraph*{We iterate until convergence.}

\section{Simulation study}

In this section we compare the different approaches with the help
of simulation. We will also verify the convergence of $\hat{\boldsymbol{\mu}}^{un}$
to the sparsest solution as claimed in the previous section. The performance
of the new proposed algorithm will be studied using simulation technique
where we will investigate to what extent we can reduce the dimension
of the observed data using the proposed approach in order to have
a fair reconstruction of the parameter.

\subsection{Convergence of the Unrestricted EM estimate:}

Here we see that in the unrestricted EM algorithm the EM estimate
of $\boldsymbol{\mu}$ converge to the sparsest solution of equation
$\eqref{eq:myequation1}$ if we take our initial estimate as $\mathbf{0}$
(or very close to $\mathbf{0}$). We take different initial estimates
of $\boldsymbol{\mu}$ randomly and check the $L_{1}$ norm of the
final estimates $\hat{\boldsymbol{\mu}}^{un}$ in each case. For demonstration
we work with $n=4$. We find that we reach the minimum norm solution
if the initial estimate of $\boldsymbol{\mu}$ is taken close to $\mathbf{0}$.\medskip{}

\medskip{}

\qquad{}\qquad{}\qquad{}%
\begin{tabular}{|c|c|}
\hline 
Initial estimate $\hat{\boldsymbol{\mu}}^{(1)}$ & $L_{1}$ norm of $\hat{\boldsymbol{\mu}}^{un}$ \tabularnewline
\hline 
\hline 
(0.0001,0.0001,0.0001,0.0001) & 10.5667\tabularnewline
\hline 
(12.52,22.76,35.98,67.72) & 38.9358\tabularnewline
\hline 
(10.5,11.25,25.62,19.74) & 27.8503\tabularnewline
\hline 
\end{tabular}

\medskip{}

\medskip{}

\subsection{Comparison of Approaches:}

Next we compare the accuracy of the different approaches discussed
in the paper. From \textbf{Theorem 2} we find that the accuracy of
the reconstructed signal is shown by $\eqref{eq:equation3.1}$ . Hence
we take $\parallel\mathbf{x-\widehat{x}}\parallel_{l_{2}}$as measure
of closeness between the original and the reconstructed signal. We
note that there is difference in the setup of the data in the approaches
$\eqref{sec:A-Naive-Approach}$. The conventional approach reconstruct
the signal $\mathbf{x}$ whereas the new approaches reconstruct what
is called true signal (free from noise) $\boldsymbol{\mu}$. Hence
for comparison we reconstruct signals from same population using conventional
approach and average out the residuals to remove the effect of the
noise.

\paragraph*{For the comparison of approaches we adopted the following technique:}
\begin{itemize}
\item We set the actual number of observations $n$ and the observed number
of observations $m$. $k$, the maximum number of nonzero components
in $\boldsymbol{\mu}$, is taken to be equal to $m$ (maximum possible
value), that is, we do not use any prior information about the number
of nonzero components in $\boldsymbol{\mu}$.
\end{itemize}
\smallskip{}

\begin{itemize}
\item We fix a $\boldsymbol{\mu}$ such that its first $4$ components are
$5$ and the rest are zero.
\end{itemize}
\smallskip{}

\begin{itemize}
\item We start with a value of $\sigma$ between $0.1$ and $1$.
\end{itemize}
\smallskip{}

\begin{itemize}
\item \textbf{Assessing Conventional Approach:} We generate data $\mathbf{x}$
from $N_{n}(\boldsymbol{\mu},\sigma^{2}I_{n})$ and reconstruct $\widehat{\mathbf{x}}$
using $\eqref{eq:equation3.3}$ from the conventional approach and
find $\parallel\mathbf{x-\widehat{x}}\parallel_{l_{2}}$. This process
is repeated 1000 times to find the residuals in each case and then
we compute the mean residual $\frac{1}{1000}{\displaystyle {\displaystyle \sum_{i=1}^{1000}}\parallel\mathbf{x_{i}-\widehat{x}_{i}}\parallel_{l_{2}}}$
to remove the effect of randomness and get a measure of closeness
among the original and reconstructed $\boldsymbol{\mu}$.
\end{itemize}
\smallskip{}

\begin{itemize}
\item \textbf{Assessing New Approaches:} We again generate data $\mathbf{x}$
from $N_{n}(\boldsymbol{\mu},\sigma^{2}I_{n})$ .We apply the naive
approach (wherever possible) and the new approach to reconstruct $\boldsymbol{\mu}$
and find $\parallel\boldsymbol{\mu}-\hat{\boldsymbol{\mu}}\parallel_{l_{2}}$
as a measure of closeness between the original and estimated values.
\end{itemize}
\smallskip{}

\begin{itemize}
\item For each value of $\sigma$ in we repeat the process of assessing
the conventional and new approaches $10$ times each to get the average
residual and standard error of the residuals for each of the conventional
and the proposed algorithms.
\end{itemize}
\smallskip{}

\begin{itemize}
\item We repeat the above procedures for different values of $\sigma$ in
$[0.1,1.0]$ and plot the mean residuals along with the standard error
bars.
\end{itemize}
\smallskip{}

For small values of $n$ we plot the average residuals for the three
approaches discussed earlier. 

\qquad{}\qquad{}\qquad{}\includegraphics[width=8cm,height=8cm]{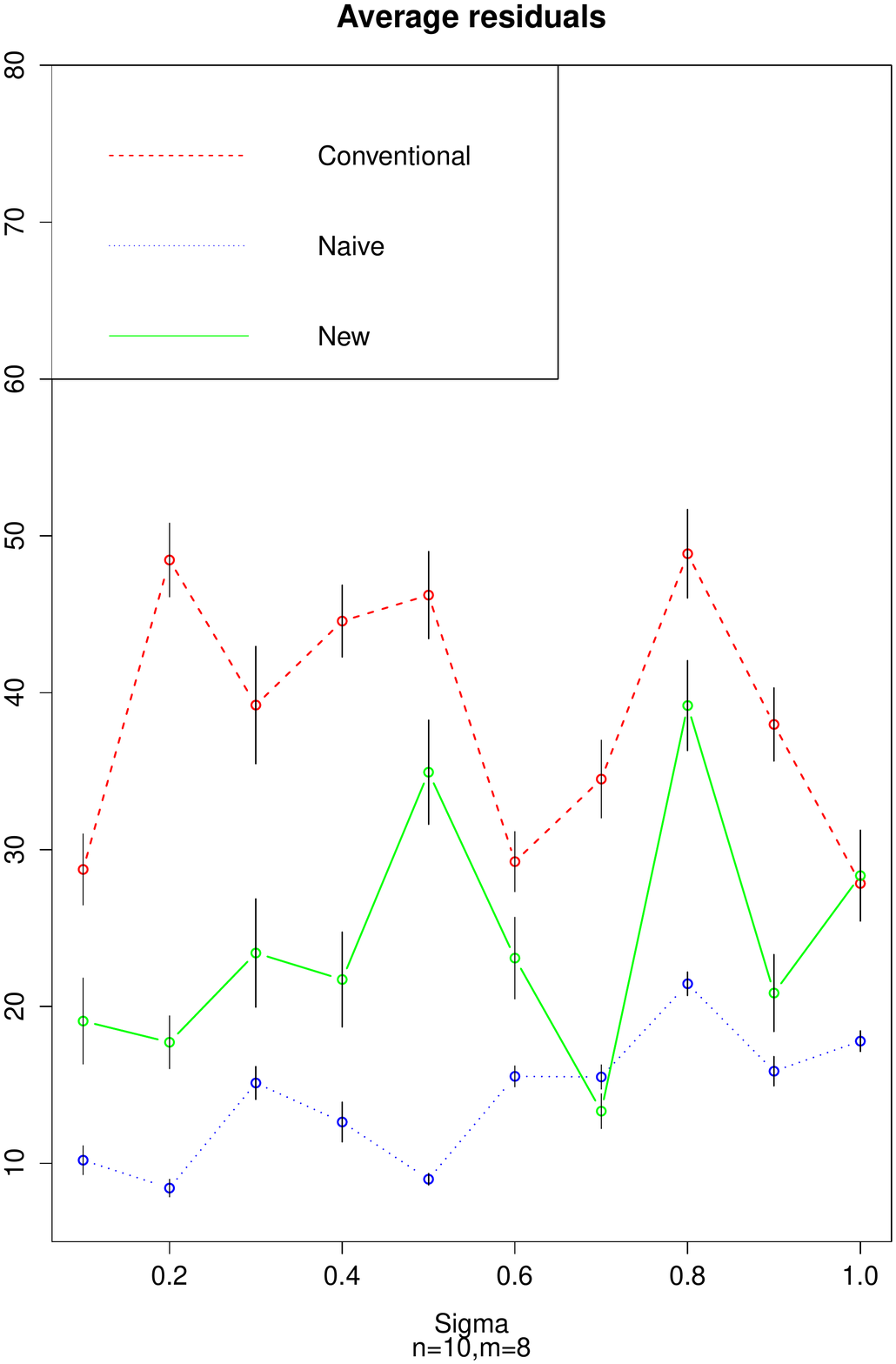}

\paragraph*{For $n=10$ we find that the naive approach works uniformly best
for different values of $\sigma$.Thus it would have been nice if
we can apply this naive approach for all values of $n$ , but unfortunately
due to the inapplicability of this procedure we turn our attention
towards the new approach. }

\paragraph*{For moderate to large values of $n$ we cannot plot the residuals
of the naive approach as it is computationally impossible. Also the
comparison between the new and the conventional approach cannot be
performed for very large values of $n$ because of computational time.
We find that the new approach works uniformly better for different
values of $\sigma$ for both $n=50$ and $n=100$.\medskip{}
}

\medskip{}

\paragraph*{\protect\includegraphics[width=6cm,height=7.5cm]{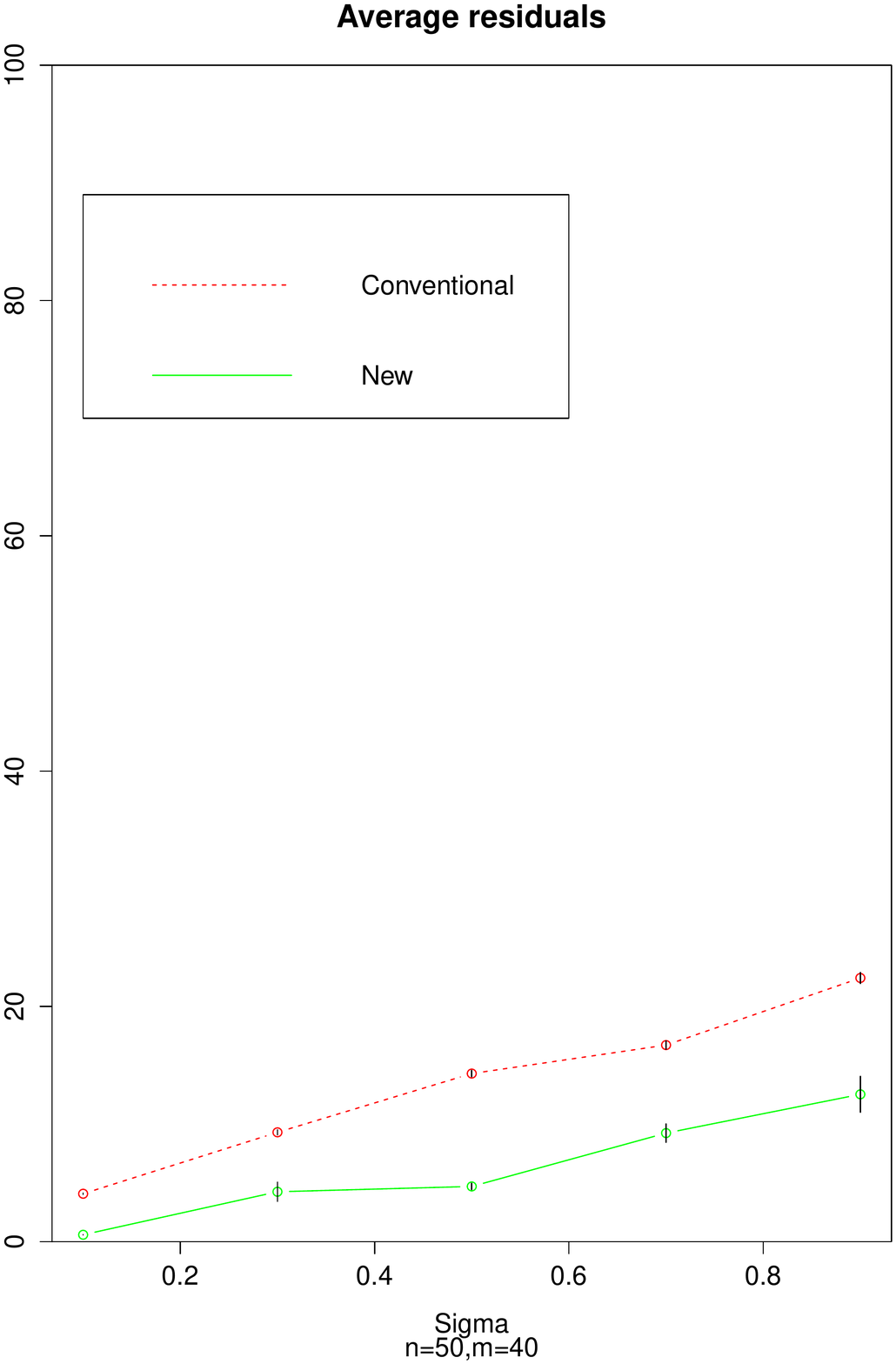}\textmd{\qquad{}}\protect\includegraphics[width=6cm,height=7.5cm]{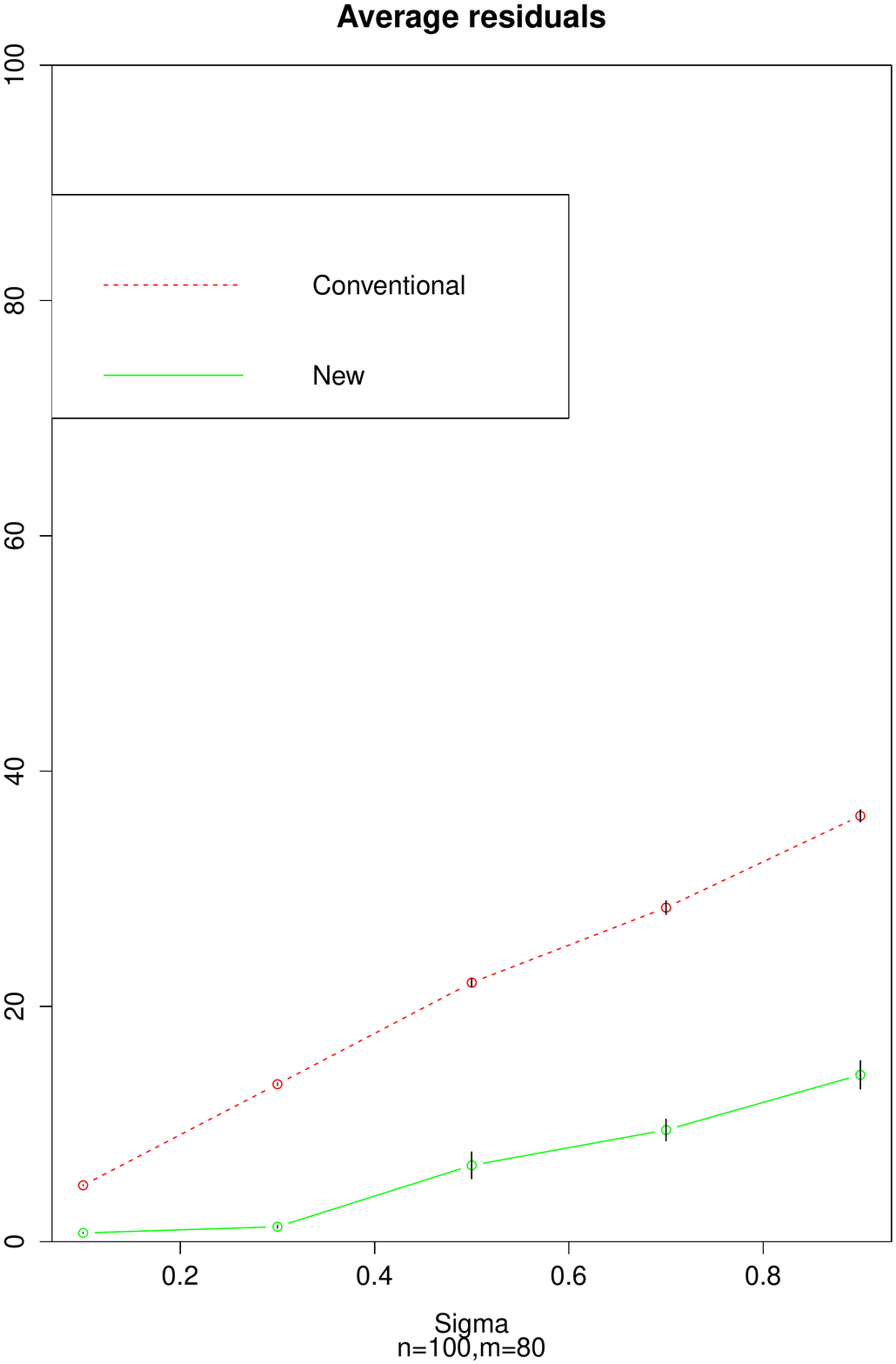}}

\subsection{Performance of the new approach:}

The value of $\frac{m}{n}$ in the above procedures is an important
point of consideration. It signifies the sampling fraction ,that is
to what extent we can reduce the dimensionality of the problem. We
fix $n=1000$ and with $\sigma=0.001$ we plot the average residuals
for varying $m$.

\paragraph*{\medskip{}
}

\qquad{}\qquad{}\qquad{}\includegraphics[width=8cm,height=6.5cm]{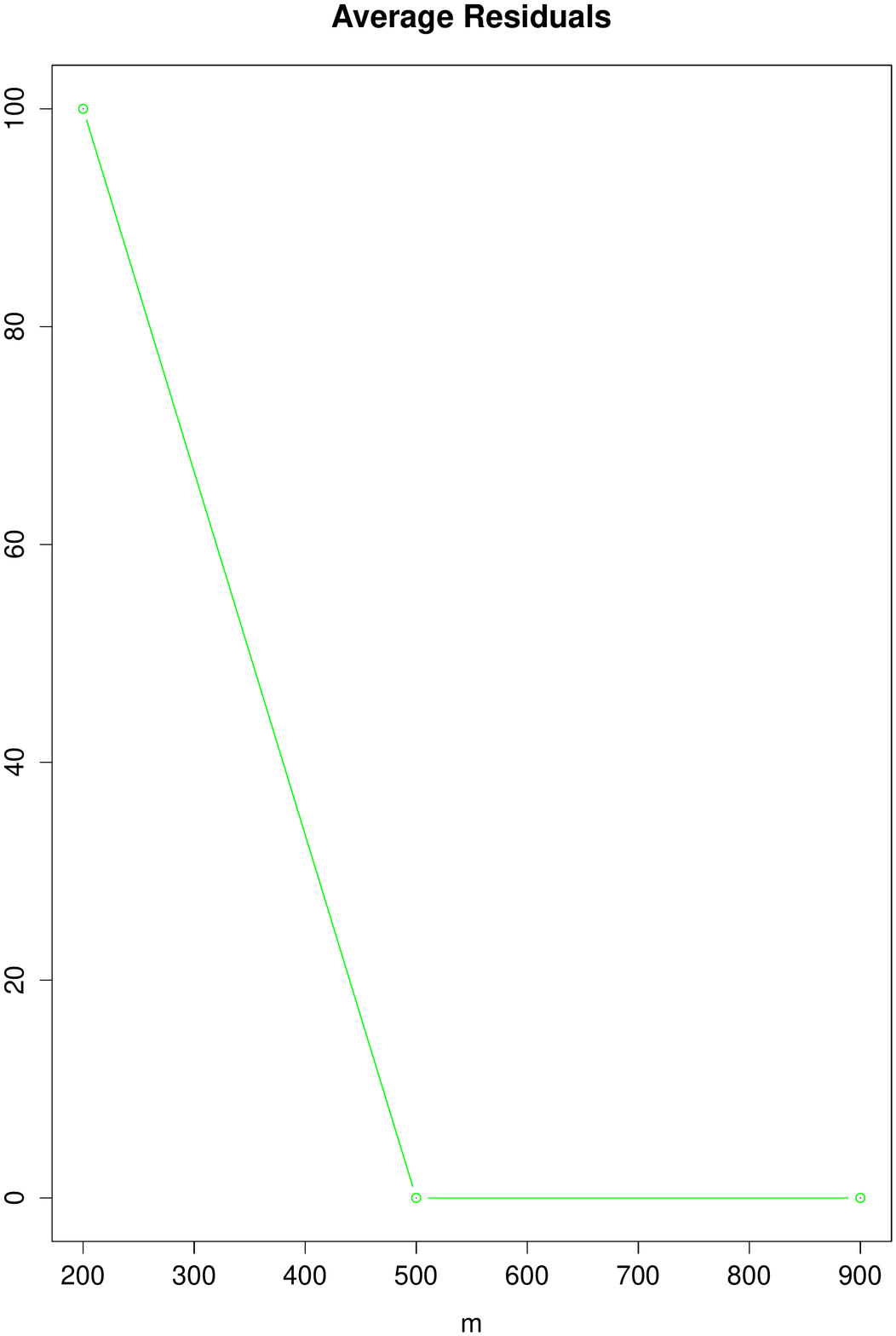}

\paragraph*{The procedure works good if we take $m=500$ ,that is at this variance
level we can afford $50\%$ dimensionality reduction.}

\paragraph*{Thus we find that the new approach works better than the conventional
method of signal reconstruction. The conventional method of reconstructing
the signal assumes the noise to be bounded with high probability and
thus fail to perform well for large error variance whereas the new
approach allows the error variance to be large enough and thus make
it applicable to other situations. Also the conventional approach
assumes that the signal is sparse and sparsity is an essential ingredient
in the reconstruction algorithm. The new proposed approach can easily
be generalized to even situations where signals need not to be sparse.
However we find that the naive approach we proposed earlier works
best if it can be implemented. For moderate to large dimensional problems
which are common in practice the new algorithm works better than the
conventional approach.}

\section{Future work}

The present paper treats observations or signals as iid samples from
a population. This can be extended assuming a non-iid setup where
the signals may be generated from a stochastic process. Further here
we work with linear combinations of all signals. A further extension
can be done where we builld the model with linear combinations of
some signals and apply it for future signals in the process.

\end{document}